# Pressure-enhanced superconductivity in cage-type quasiskutterudite $Sc_5Rh_6Sn_{18}$ single crystal

Govindaraj Lingannan[1,2], Boby Joseph[1*], Muthukumaran Sundaramoorthy [2], Chia Nung Kuo[3,4], Chin Shan Lue[3,4] and Sonachalam Arumugam[2#]

[1]Elettra-Sincrotrone Trieste S.C. p. A., S.S. 14, Km 163.5 in Area Science Park, Basovizza 34149, Italy

[2]Center for High Pressure Research, School of Physics, Bharathidasan University, Tiruchirappalli 620024, India

[3]Department of Physics, National Cheng Kung University, Tainan 70101, Taiwan

[4]Taiwan Consortium of Emergent Crystalline Materials, Ministry of Science and Technology, Taipei 10601, Taiwan

*Corresponding authors: #sarumugam1963@yahoo.com, *boby.joseph@elettra.eu

**Abstract**

$Sc_5Rh_6Sn_{18}$ with a cage-type quasiskutterudite crystal lattice and type II superconductivity, with superconducting transition temperature $T_c$ = 4.99 K, was investigated under hydrostatic high-pressure (HP) using electrical transport, synchrotron X-ray diffraction (XRD) and Raman spectroscopy. Our data show that HP enhance the metallic nature and $T_c$ of the system. $T_c$ is found to show a continuous increase reaching to 5.24 K at 2.5 GPa. Athough the system is metallic in nature, Raman spectroscopy investigations at ambient pressure revealed the presence of three weak modes at 165.97, 219.86 and 230.35 cm$^{-1}$, mostly related to the rattling atom Sc. The HP-XRD data revealed that the cage structure was stable without any structural phase transition up to ~7 GPa. The lattice parameters and volume exhibited a smooth decrease without any anomalies as a function of pressure in this pressure range. In particular, a second order Birch-Murnaghan equation of state can describe the pressure dependence of the unit cell volume well, yielding a bulk modulus of ~ 97 GPa. HP Raman investigations revealed a linear shift of all the three Raman modes to higher wavenumbers with increasing pressure up to ~8 GPa. As the pressure enhances the bond overlap, thus inducing more electronic charges into the system, HP-XRD and Raman results may indciate the possibility of obtaining higher $T_c$ with increasing pressures in this pressure range.

Supplementary material for this article is available online.

## 1. Introduction:

Caged-type compounds have got a lot of attention recently because of their several unusual features [1]. Three cage type compounds have been actively explored as "rattling-good" materials: Such as β-pyrochlore oxides [2,3,4], Ge/Si clathrates [5,6] and filled skutterudites [7,8]. A Rare earth (R) guest atom was trapped in a highly symmetric cage in these compounds, resulting in its unique behavior. Ternary stannide systems $R_5Rh_6Sn_{18}$ (R– rare-earth elements like Sc, Y, and Lu) comes under this crystal structure category. There is also the appearance of superconductivity with transition temperature ($T_c$) at 5 K for R=Sc [9], 3 K for R=Y [10] and 4 K for R= Lu [11]. R atom produce the strong interplay between the quadrupolar moment and superconductivity in $R_5Rh_6Sn_{18}$ [12,13]. The above three ternary stannides, $R_5Rh_6Sn_{18}$, (R = Sc, Y, and Lu) are found to exhibit nodeless superconducting gap [11] and s-wave gap function [10,14,15]. Isotropic superconducting gap and conventional BCS type superconductivity were observed in both $Sc_5Rh_6Sn_{18}$ [16] and $Lu_5Rh_6Sn_{18}$ [17]. This result is in contrast with $Y_5Rh_6Sn_{18}$, which show highly anisotropic gap [17,18]. This implies that the superconducting gap structure is highly influenced by the R atom. However, such a predominant rare-earth atom dependence on the SC gap structure is yet to be well understood. Estimated zero-temperature upper critical field [$H_{c2}(0)$] were 7.2 T for $Sc_5Rh_6Sn_{18}$ [9], 4.3 T for $Y_5Rh_6Sn_{18}$ and 5.2 T for $Lu_5Rh_6Sn_{18}$ [11]. The R dependence of the superconducting properties seem to suggest a good correlation between the structural details and superconductivity in these compounds. In fact, cage structured compounds are found to be interesting from a superconductivity point of view. There are several examples available, see for example Refs. [19-22]. Interestingly, the buky ball derived $K_3C_{60}$ have also got attention for the possible light inducted superconductivity [23].

The $R_5Rh_6Sn_{18}$ system forms a tetragonal crystal structure [see Fig. S1]. with a $I4_1/acd$ space group [15] where two different symmetry positions are occupied by R atoms [24]. Two cubic blocks form a distinct superlattice along the c-axis in the tetragonal unit cell [9,24], [see Fig. S2]. Atomic disorder was found to enhance the $T_c$ of $Ca_x$ doped $Y_{(1-x)}Ca_xRh_6Sn_{18}$ from 3.08 K for x=0 to 3.10 K for x=0.5 [25]. Similarly, quasiskutterudite-related compounds such as $La_3Rh_4Sn_{13}$ and $Ca_3Rh_4Sn_{13}$ shows a superconductivity, with $T_c$ respectively 2.12 K and at 4.71 K. In $Ca_3Rh_4Sn_{13}$ circumstance doping decreases the $T_c$ to 4.53 K for $Ca_{2.8}Ce_{0.2}Rh_4Sn_{13}$ and 4.50 K for $Ca_{2.4}La_{0.6}Rh_4Sn_{13}$ [25]. In this context negative pressure coefficient of $T_c$ was observed in $La_3Ru_4Sn_{13}$ and $La_3Ru_{0.5}Co_{3.5}Sn_{13}$ with the rate of respectively −0.03 K/GPa and −0.12 K/GPa in contrast to the positive pressure coefficient ∼ 0.03 K/GPa for $La_3Co_4Sn_{13}$ [26]. All these points to a delicate interplay between the structural parameters and the superconductivity in these systems, thus motivating further systematic HP studies. Here we present the HP studies on quasi-skutterudite cage type $Sc_5Rh_6Sn_{18}$ compound. It is recalled that Sc atoms have a relatively small ionic radius (Shannon's effective ionic radii [pm] for $Sc^{3+}$ is 74.5) and due to the rattling motion of the Sc atom, there is a strong electron-phonon interaction [9] in this system.

2. **Experimental methods**

Single crystals of $Sc_5Rh_6Sn_{18}$ was grown by a conventional Sn-flux method. Ref. [11] contains a detailed description of the synthesis procedures. Briefly, high-purity Sc (99.9%), Rh (99.95%), and Sn (99.999%) ingots were obtained from commercial supplier Ultimate Materials Technology and mixed in the ratio of 1:1.2:20 and sealed in an evacuated quartz tube with a flat bottom. The mixed elements were heated to 1050 °C for 12 hours, dwelled for 20 hours, then cooled down to 600 °C in 150 hours. Excess Sn flux was separated from the crystals by centrifugation and by etching in dilute HCl. High quality large crystals could be obtained by this procedure [see Fig. S2(c)].

Ambient and pressure dependent x-ray powder diffraction (HP-XRD) data were collected at the Xpress beamline [27] of the Elettra Synchrotron Center, Trieste, Italy at room temperature. A membrane-driven symmetric diamond anvil cell (DAC) with culet size 400 µm and PACE-5000 based automatic membrane drive was used. Sample chamber for the HP-XRD were prepared by intending a 200 µm thickness 301 stainless steel metal foil to 50 µm and drilling a 160 µm through hole in the center. Pressure transmitting medium used was 4:1 methanol ethanol mixture. In-situ ruby florescence technique was used to monitor the pressure inside the DAC by including ~10 µm ruby chips in the sample chamber along with the sample. A monochromatic circular beam with a wavelength of 0.4957 Å and a cross-sectional diameter of 40 µm was used to collect high-pressure diffraction data. A PILATUS3S-6M large-area detector was used for diffraction data collection. FIT2D software was employed for converting the diffraction image files from the PILATUS3S-6M to 2-theta vs intensity plots. The Rietveld refinement using the GSAS-II suite was carried out to obtain the structural parameters as function of pressure. Atomic positions and thermal parameters were refined only for the ambient condition diffraction data and were kept fixed for the analysis of the high-pressure data. Ambient pressure structural details at room temperature are reported in the table S1. These are in good agreement with those reported recently [28].

The Raman scattering measurements were performed in a customized Renishaw Raman confocal microscope available at the Xpress beamline using 532 nm (green) diode laser with 2400 l/mm gratings at ambient and various hydrostatic pressures (up to 7.5 GPa). Membrane diamond anvil cell (DAC) pressure cell (Betsa, France) and 4:1 methanol ethanol mixture was used as pressure transmitting. In-situ ruby florescence approach was employed to monitor the pressure inside the DAC. Our sample is pure metallic, and observed three extremely weak Raman modes in it. For signal reliability, ten runs were performed in each pressure at various points on the sample. After applying the pressure the pressure cell was held at each pressure for enough time for further ensuring the hydrostatic conditions. To obtain accurate results, the measured data sets were averaged after taking care of the background properly.

The electrical resistivity of a bar-shaped single crystal sample was measured in a cryogen-free closed-cycle refrigerator (CCR-VTI) system at temperatures ranging from 4 to 300 K using a standard linear four-probe DC resistivity technique. The electrical contacts on the top of the sample were made with silver paste (4929N) and copper wire (0.05 mm). A clamp-type hybrid double-cylinder piston pressure cell was used to measure hydrostatic high-pressure electrical resistivity, pressure cell inner cylinder and outer cylinder was made up of NiCrAl and BeCu. Daphne oil 7474 was used as a pressure transmitting medium for superior hydrostaticity below 3.7 GPa [29] and the actual pressure inside the cell was calculated using a calibration curve obtained previously from bismuth fixed-pressure points. A 20-ton hydraulic press (Riken Kiki, Japan) was used to provide pressure to the pressure cell, which is then clamped with the desired pressure [30].

Bismuth exhibits structural phase transitions I–II at 2.55 GPa and II–III at 2.7 GPa at room temperature. Due to these structural transitions, two distinct jumps in resistance were observed at applied pressures of 17.6 and 19 MPa. Similar to an earlier report by Honda *et al.,* [31] a linear calibration curve was drawn by interpolating these applied pressures versus actual pressure data.

## 3. Results and discussion

Figure 1 shows the temperature dependence of electrical resistivity measurements under various hydrostatic pressure. Figure S3 shows the temperature dependent of resistivity measurement at ambient pressure. At ambient pressure our sample showed a good metallic behavior down to 100 K. Fig. 1 (b) shows the magnified view of resistivity from 300 K to 5 K. Resistivity decreases with decreasing temperature down to 100 K. With further decreasing temperature the rate of decrease of resistivity is observed to be lower. This is an indication of a bad metallic behavior. In fact, Feig *et al.,* observed an anisotropic bad metallic behavior parallel to c axis. This is ascribed to the strong structural disorder and low charge carrier concentration [16,28]. Our data is similar to that reported by Feig et al.,[16]. As can be readily appreciated from the inset in Fig. S3, our data shows a sharp fall of resistivity at 5.09 K which reaches zero at 4.93 K revealing a clear superconducting transition. $T_c$ is calculated from the derivative plot, as illustrated in the inset (Fig. S3), whose value in this case is 4.99 K. The presence of a sharp superconducting transition again underlines the excellent quality of our $Sc_5Rh_6Sn_{18}$ sample. Observed $T_c$ is good agreement with the previous reports [9,14,16,28]. The data shown in Fig. 1 reveals a pressure enhanced metallic nature up to the $T_c$ until 2.5 GPa and the bad metallic behavior is gradually being suppressed by the application of pressure. Inset Fig. 1(a) shows the magnified view of superconducting transition temperature region and it reveals positive pressure coefficient of $T_c$. In particular, $T_c$ at 0 GPa is 4.99 K which increase to 5.24 K at 2.5 GPa with an increment rate $dT_c/dP = 0.1$ K/GPa. For completeness we present also the $T_c^{onset}$ and $T_c^{zero}$ values in Fig. S4.

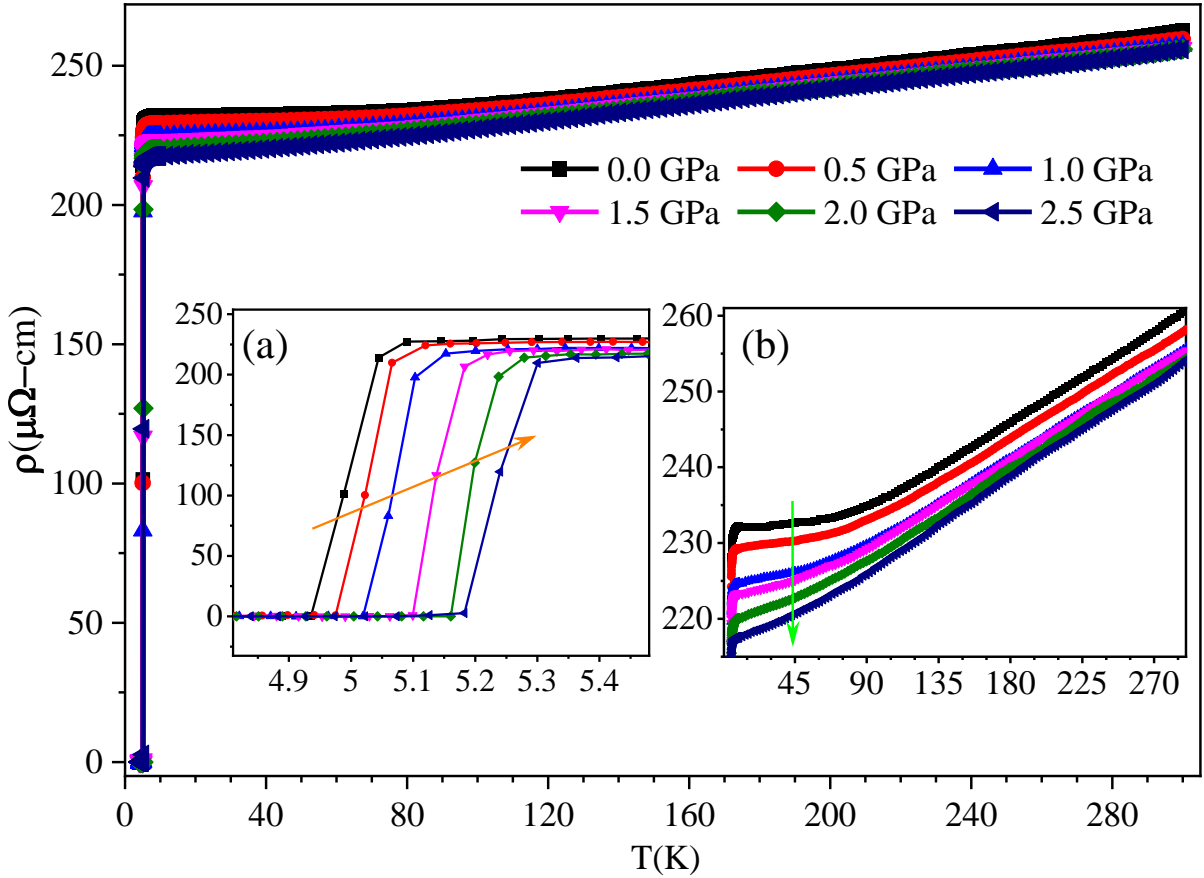

Figure 1: Temperature dependence of electrical resistivity measurements under various hydrostatic pressure; Inset (a) shows an enlarged view of the low-temperature superconducting region. Inset (b) shows a zoom view of resistivity above the superconducting transition.

Both the enhancement of the metallicity and the $T_c$ by the application HP are due to the enhancement of electronic density of states at the Fermi level [$N(E_F)$]. A similar behavior is observed in $Ba_6Ge_{25}$ cage compound [32]. However, there is a negative pressure resistivity effect in this type of spin-fluctuation system which is due to the hybridization with conduction electron states produces enhanced delocalization of $f$ or $d$ states [33]. On the other hand, chemical pressure is found to enhance the superconductivity in isostructural compounds $Y_5Rh_6Sn_{18}$ [34] and $Lu_5Rh_6Sn_{18}$ [35]. Similar $La_3Co_4Sn_{13}$ skutterudite showed a positive pressure coefficient of $T_c$ with the rate of ~0.03 K/GPa. Presence of a subtle structural distortion ($T_D$) at 140 K was observed in this compound, which was attributed to the deformation of the $Sn_{12}$ cages and consequent Fermi-surface reconstruction. Further, $T_D$ is found to shift towards low temperature by the application of pressure with a corresponding increase also in the $T_c$. A change in the slope of $1/\rho$ vs T was used to define $T_D$. [36]. Extracted $T_D$ values for our $Sc_5Rh_6Sn_{18}$ sample are presented in Fig. S5. In this context we recall that most of the other skutterudite cage compounds showed a negative pressure coefficient of $T_c$ [25,26,33].

Figure 2 (a) shows measured XRD patterns of $Sc_5Rh_6Sn_{18}$ up to 6.89 GPa of pressure using 4:1 methanol-ethanol as pressure transmitting medium. Only selected pressure patterns are showed and the patterns were moved vertically, both to have a better presentation of the data. The pattern is similar across all pressures, with the exception of a peaks move towards higher angle (2θ) values as pressure rises (see also Fig. S6), these are indicating that no symmetry changing structural phase transitions exist in this pressure range. The ambient pressure refinement results show that the synthesized sample has a tetragonal crystal structure with a I4$_1$/*acd* space group. The lattice parameters are found to be a = 13.57493 Å and c= 27.10011 Å. Table S1 provides the structural details as obtained from the Rietveld refinement. These results are in good agreement with the previous reports [16,28]. Synchrotron diffraction rule out the possibility of any impurity phases in the sample. Fig. 2(b) shows the measured high pressure XRD pattern together with the Rietveld refinement results at two selected pressures. Figure S7 shows the evolution of lattice parameters as a function of pressure. In-plane lattice parameter is found to decrease at a rate of -0.037 Å/GPa. The rate of the out of plane lattice parameter is -0.087 Å/GPa.

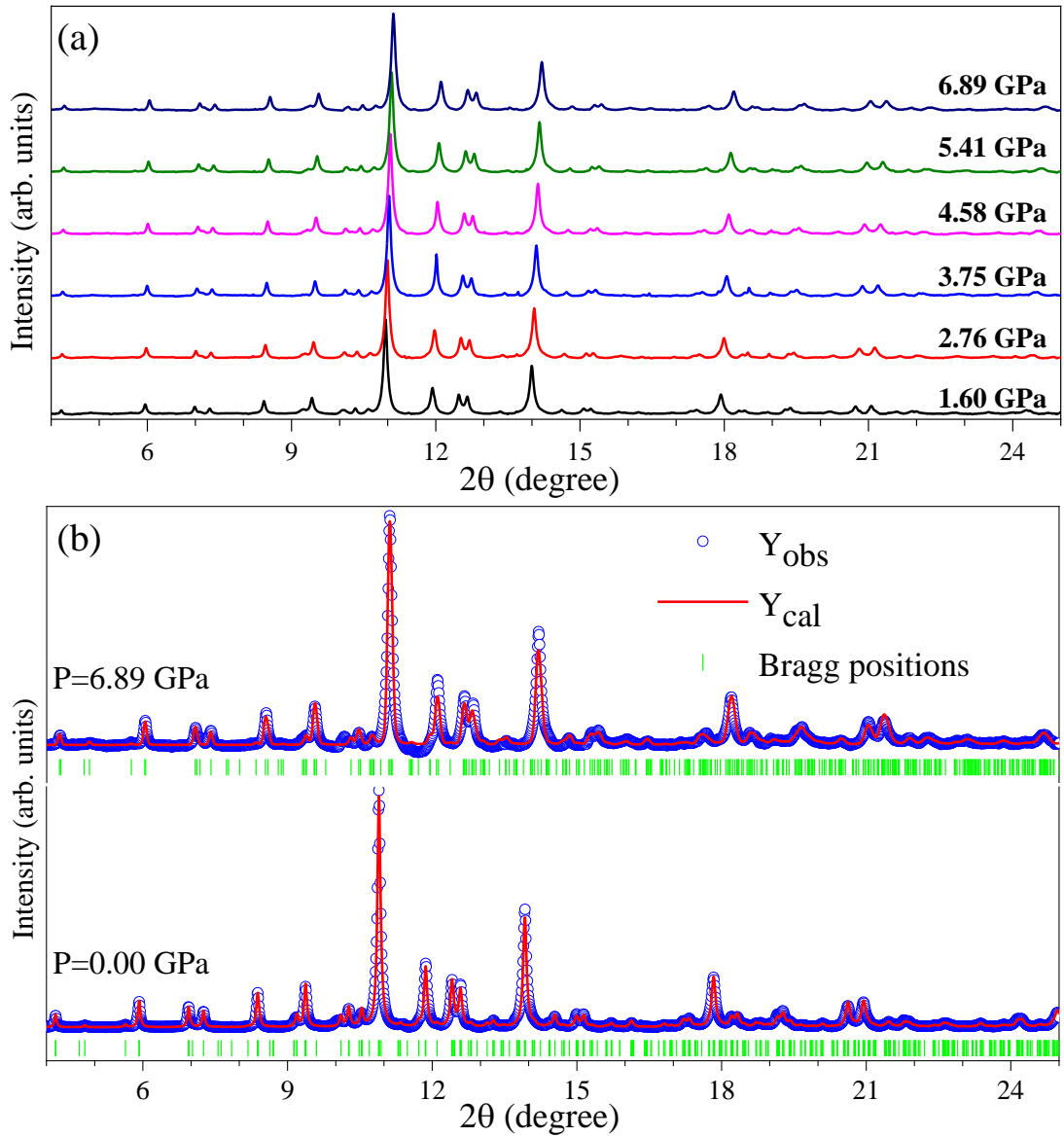

Figure 2: (a) Measured XRD patterns of $Sc_5Rh_6Sn_{18}$ at different selected pressures up to 6.89 GPa. The patterns were stacked vertically with an off-set for clarity in presentation. (b) XRD pattern together with the Rietveld refinement results at ambient pressure (indicated as P = 0.00 GPa) and P = 6.89 GPa.

Fig. 3(a) shows the pressure evolution of the observed Raman modes in $Sc_5Rh_6Sn_{18}$. Such weak Raman modes in the low wavenumber regions are reported for similar skutterudites [37,38,39]. These modes are attributed the rattling atom inside the cage [37,40, 41]. Based on these data, we attribute the observed three weak modes to be due to the rattling atom Sc in $Sc_5Rh_6Sn_{18}$. We note that to the best of our knowledge this is first Raman report on such superconducting tetragonal skutterudites systems. Raman mode analysis using peak fitting for the ambient and 7.50 GPa are shown in panels (b) and (c) of Fig. 3. This analysis yield 165.97, 219.86 and 230.35 cm$^{-1}$ as the mode frequencies at ambient pressure. Systematic theoretical analysis is needed for the obtaining the mode characterization which are being planned for a future work involving similar tetragonal skutterudites with other rare-earth atoms.

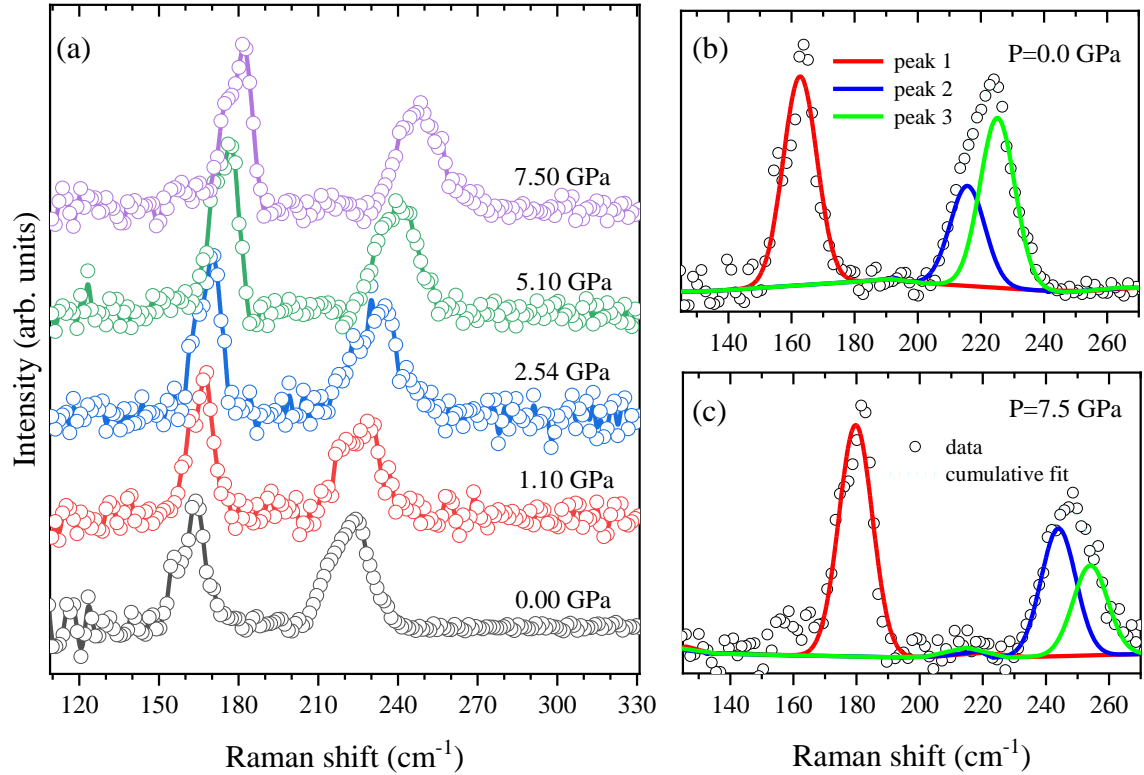

Figure 3: (a) Measured Ramandata of $Sc_5Rh_6Sn_{18}$ at five different pressures. Numbers close to the spectra are the pressure values in GPa. Peak fitting results of the ambient (b) and 7.5 GPa (c) are also shown.

Fig. 4(a) shows the evaluation of $T_c$ under pressure upto 2.5 GPa, the increment rate of $dT_c/dP$ = 0.1 K/GPa, it is comparable good enhancement compare with $La_3Co_4Sn_{13}$ compound [36]. Evolution of unit cell volume as a function of pressure is presented in Fig. 4(b). Solid red line here is the result of a second-order Birch-Murnaghan equation of state (EoS) fit to the data. As can be appreciated from the figure, the variation in the unit-cell volume in the investigated pressure range is well characterized by a second-order Birch-Murnaghan EoS. Analysis provided the bulk modulus ($K_0$) to be 99.27 GPa, with zero-pressure unit-cell volume ($V_0$) 4988.95 Å$^3$. The pressure dependence of the three Raman modes are shown in Fig. 4(c). A linear increase in the mode wavenumbers is observed with pressure.

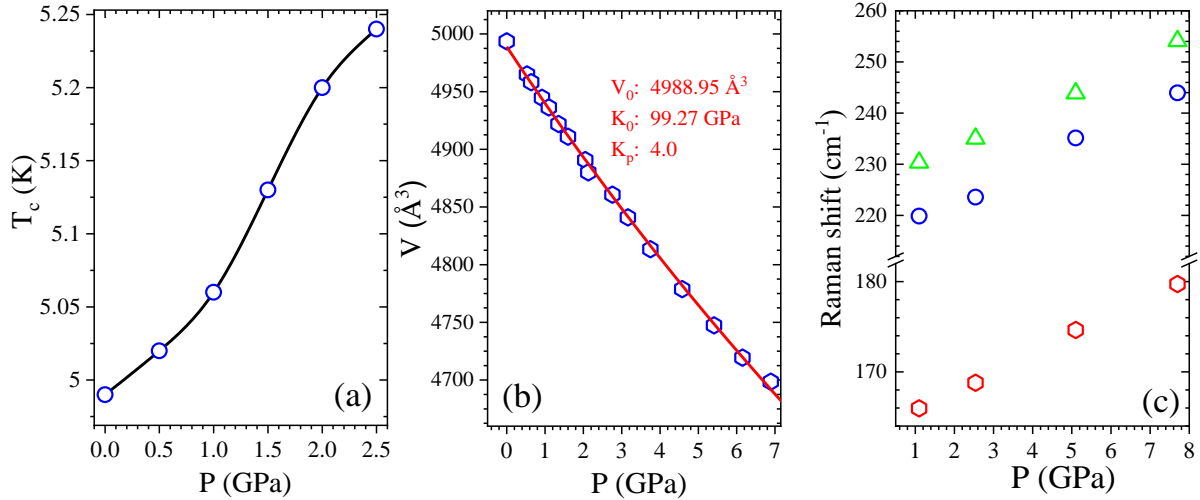

Figure 4: (a) $T_c$ as a function of pressure. (b) Unit cell volume up to ~7 GPa as a function of pressure: symbol - experiment, solid-line - fit of a second-order Birch-Murnaghan equation of state. Pressure dependence of the observed three Raman modes (c).

## 4. CONCLUSIONS

We have presented the results of systematic high-pressure (HP) resistivity measurements, HP Raman and HP synchrotron XRD of single crystal of cage-type $Sc_5Rh_6Sn_{18}$. Ambient pressure XRD conforms our grown single crystal is single phase and with no impurity phases. At Ambient pressure, superconducting transition is observed at 4.99 K. Ambient pressure Raman spectroscopy investigations revealed the presence of three weak modes at 165.97, 219.86 and 230.35 cm$^{-1}$, mostly related to the rattling atom Sc. Application of hydrostatic pressure is found to enhance the superconducting transition temperature, which reached to 5.24 K at 2.5 GPa. Most of the skutterudite cage compounds showed a negative pressure coefficient only few shows a positive pressure coefficient such as $La_3Co_4Sn_{13}$. The improvement of superconductivity is attributed to an increase in electronic density of states at the Fermi level. We found a linear decrease of lattice parameters, and volume in high pressure XRD measurement and a linear decrease in the phonon mode frequencies from Raman meaurements, all indicating no evidence for any structural phase transition or any other structural anomalies in the studied pressure range. Our result gives the evidence of enhancement of $T_c$ in the low pressure regime up to 2.5 GPa. Based on the current HP-diffraction and Raman data, it is interesting to investigate whether this increase can be continuous even up to ~ 7 GPa.

**Acknowledgments**

The authors thank Xpress beamline of the Elettra Sincrotrone Trieste for beamtime. G. L gratefully acknowledges the receipt of a fellowship from the ICTP Programme for Training and Research in Italian Laboratories, Trieste, Italy. B. J acknowledge the availability of DACs through Xpress-PLUS internal project of Elettra Sincrotrone Trieste. A. S wishes to thank DST (ASEAN, PURSE, FIST, JSPS, MES, and SERB), UGC-DAE CSIR (Indore), MHRD-RUSA, TANSCHE (Chennai), and BRNS (Mumbai) for financial support.

**Data availability statement**

The data that support the findings of this study are available from the corresponding authors upon reasonable request.

**Declaration**

The authors declare that there are no conflicting interests involved and there is no ethical issues associated.

**Supplimentary information**

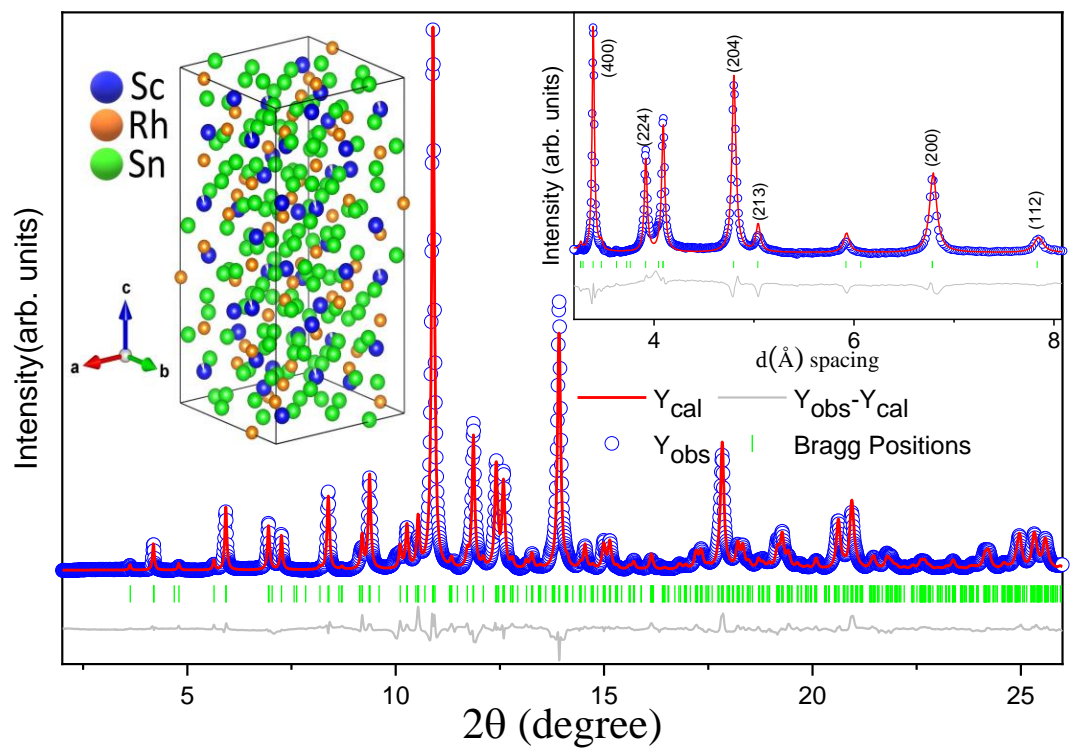

Figure S1: Measured x-ray-diffraction pattern of $Sc_5Rh_6Sn_{18}$ sample at ambient pressure and room temperature (symbols) together with the Rietveld refinement results (solid red lines). Green vertical bars below the pattern indicate the Bragg peak positions. The inset shows a structure of $Sc_5Rh_6Sn_{18}$ and a magnified view of the lower d spacing.

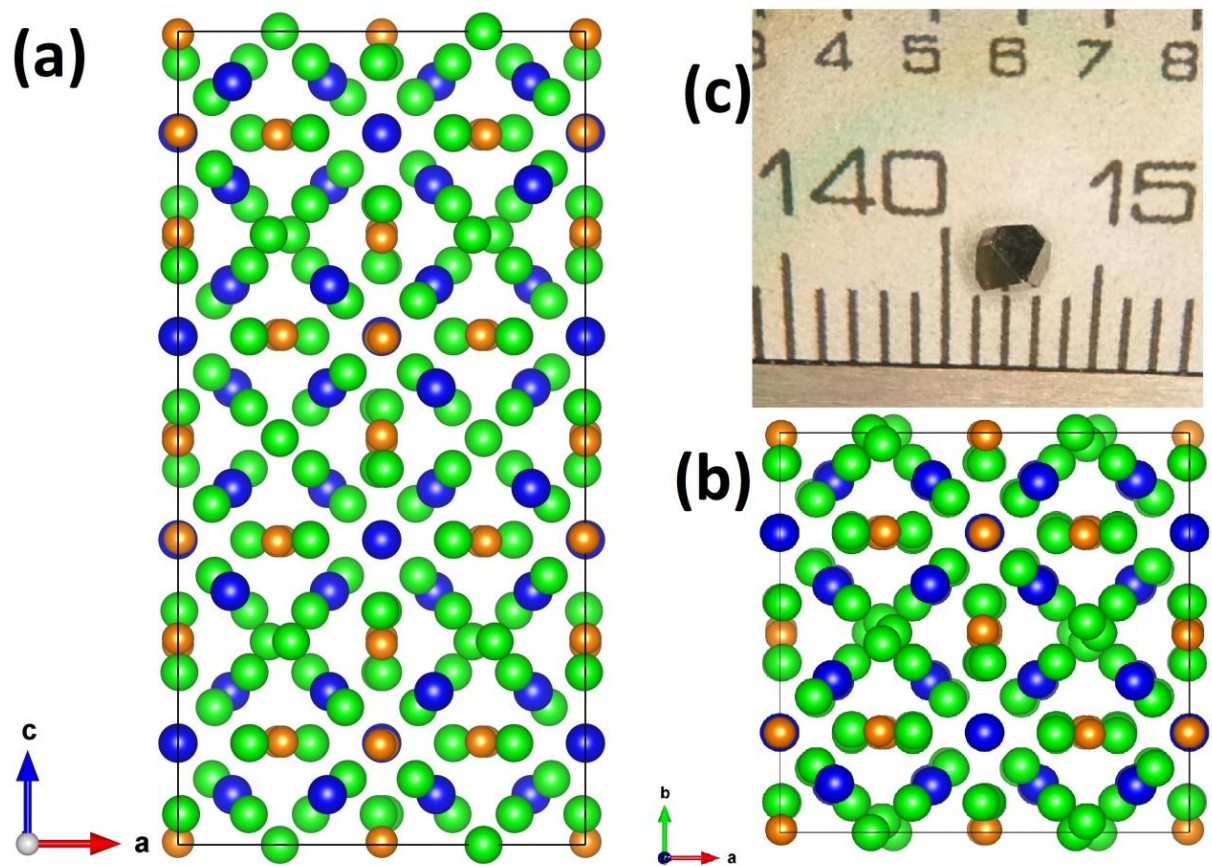

Figure S2: Structure of $Sc_5Rh_6Sn_{18}$ along ac plane (a), ab plane (b) and photo of typical single crystal (c).

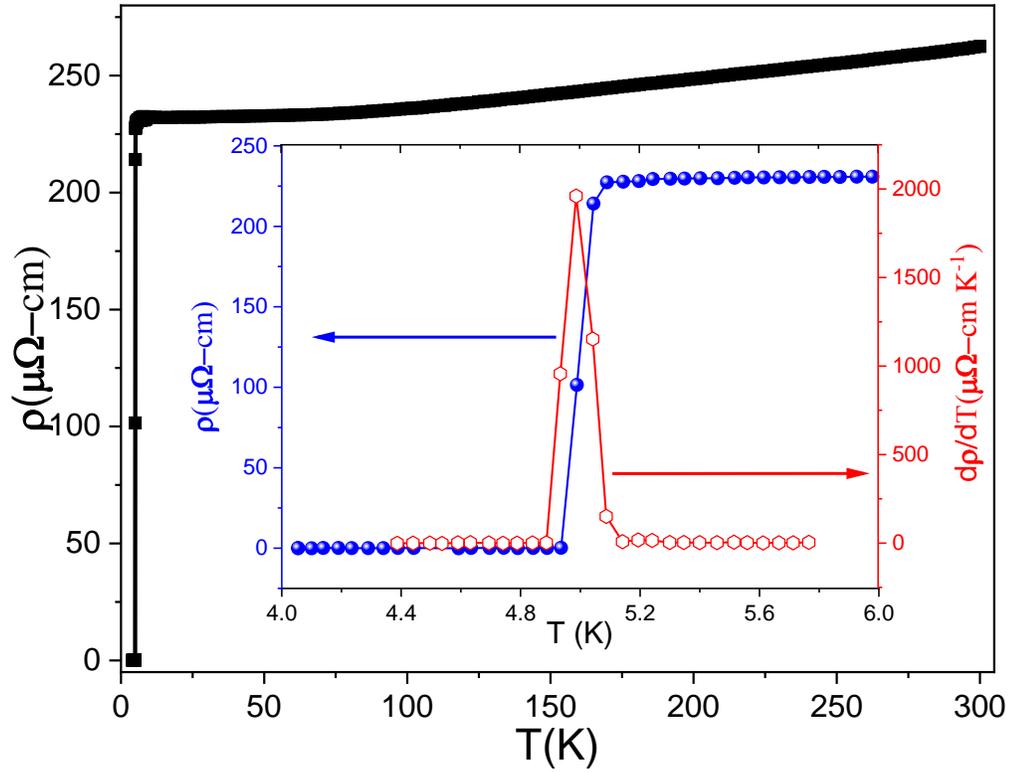

Figure S3. Temperature dependent of resistivity at ambient pressure; Inset shows an enlarged view of the low-temperature superconducting region (left) and the resistivity first derivative (right) of $Sc_5Rh_6Sn_{18}$ sample.

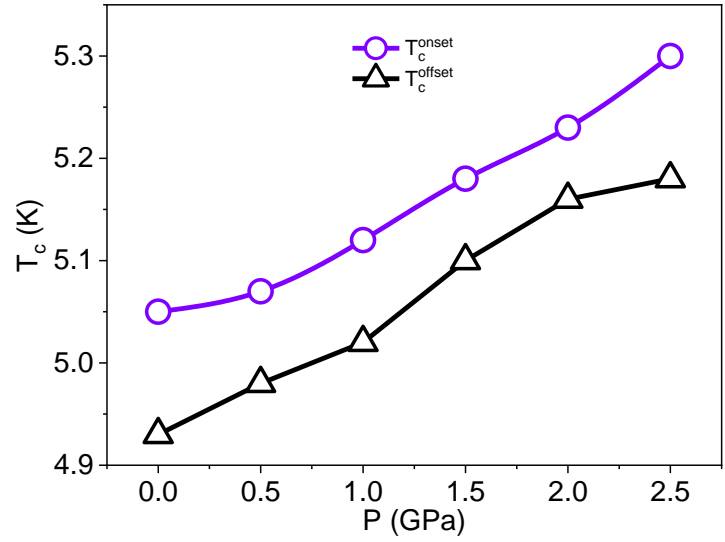

Figure S4: shows the $T_c^{onset}$ and $T_c^{zero}$ as a function of Pressure

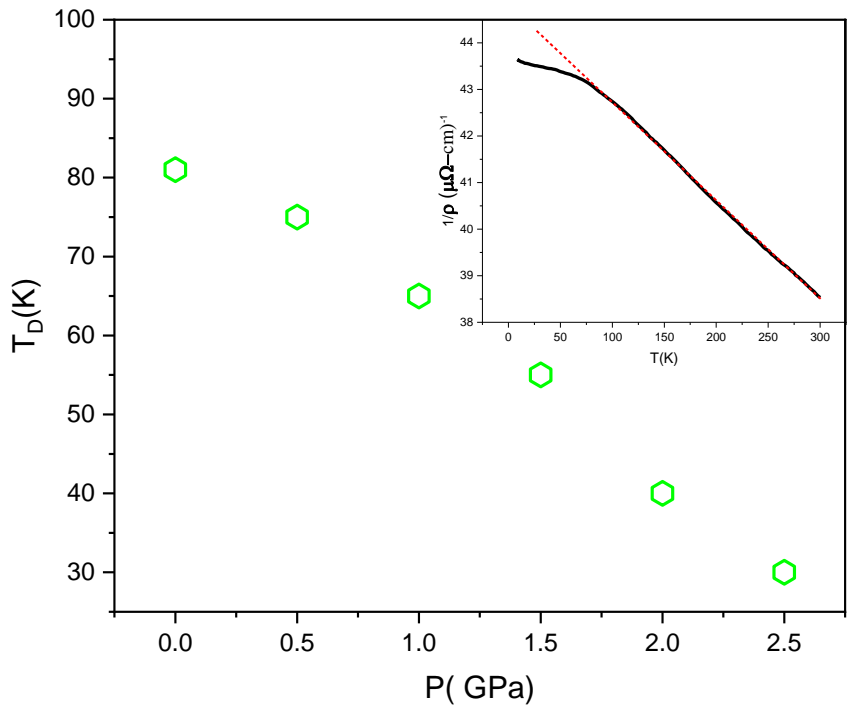

Figure S5. Structural distortion in $Sc_5Rh_6Sn_{18}$ as a function of Pressure. Inset shows the $1/\rho$ vs T

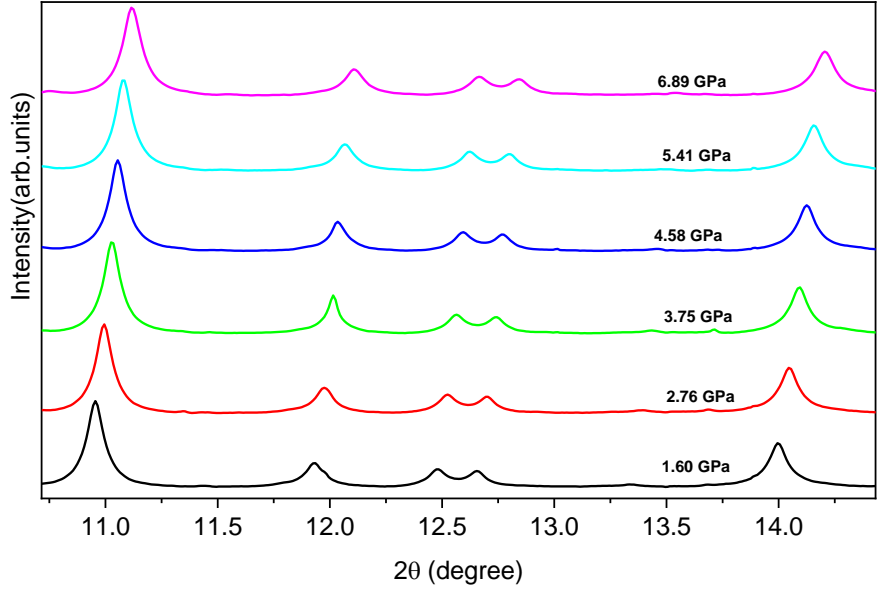

Figure S6. Magnified view of XRD patterns of $Sc_5Rh_6Sn_{18}$ up to 6.89 GPa demonstrating a smooth shift of the Bragg peaks.

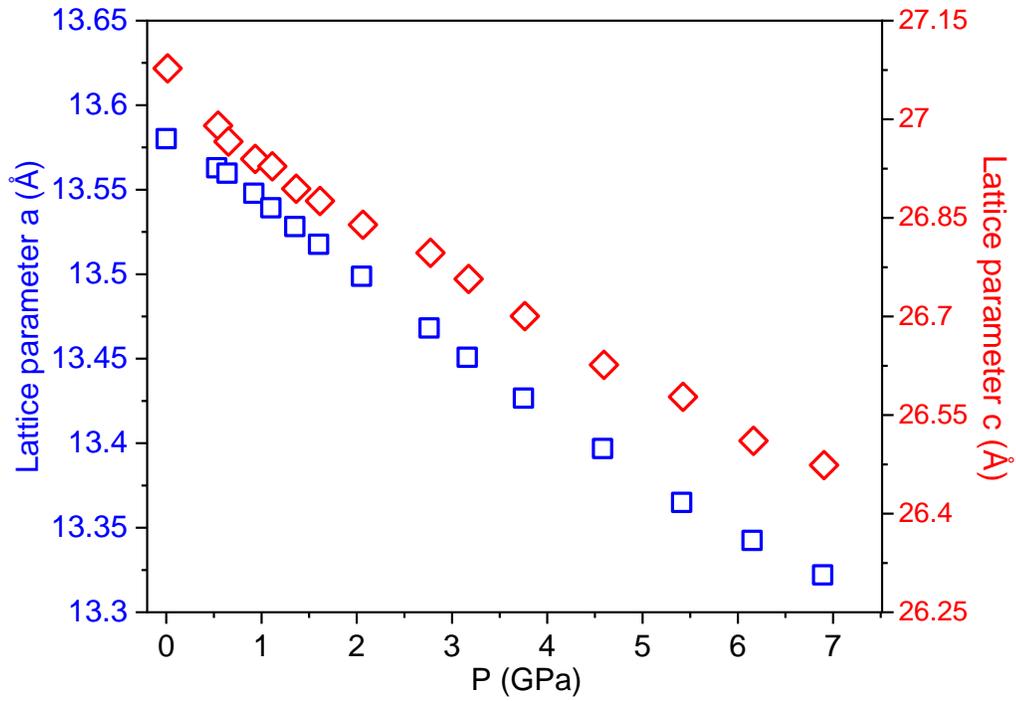

Figure S7. Lattice parameters evolution as a function of pressure

**Table S1** Refined Atomic coordinates (x, y and z), occupational parameters (fraction), Site Symmetry, multiplicity and isotropic ($U_{iso}$) thermal displacement parameters for $Sc_5Rh_6Sn_{16}$ at ambient pressure and room temperature. The lattice parameters are a = 13.57493 Å and c = 27.10011 Å.

| Atom | Type | x | y | z | fraction | Site Sym | multiplicity | U_iso |
|---|---|---|---|---|---|---|---|---|
| Sc1 | Sc | 0 | 0.25 | 0.125 | 0.92 | 222(z) | 8 | 0.0312 |
| Sc2 | Sc | 0.12804 | 0.11662 | 0.31168 | 1 | 1 | 32 | 0.0123 |
| Rh1 | Rh | 0 | 0.25 | 0.25363 | 1 | 2(z) | 16 | 0.0121 |
| Rh2 | Rh | 0.25657 | 0.25368 | 0.12683 | 1 | 1 | 32 | 0.0089 |
| Sn1 | Sn | 0.27743 | 0 | 0.25 | 1 | 2(x) | 16 | 0.017 |
| Sn2 | Sn | 0.32336 | 0.57336 | 0.125 | 1 | 2(xy) | 16 | 0.0177 |
| Sn3 | Sn | 0.17455 | 0.42455 | 0.125 | 1 | 2(xy) | 16 | 0.0253 |
| Sn4 | Sn | 0.0831 | 0.1602 | 0.41915 | 1 | 1 | 32 | 0.0344 |
| Sn5 | Sn | 0.17376 | 0.25344 | 0.03772 | 1 | 1 | 32 | 0.0197 |
| Sn6 | Sn | 0.00749 | 0.07459 | 0.03778 | 1 | 1 | 32 | 0.0185 |